\documentclass[aps,prl,reprint,showpacs,showkeys,superscriptaddress,preprintnumbers]{revtex4-1}
\usepackage{amsmath,amssymb,bm,mathrsfs}
\usepackage{srcltx,hyperref}
\usepackage{dsfont}
\usepackage{epsfig}
\usepackage{slashed}
\usepackage{bbold}
\usepackage{psfrag}
\usepackage{color}
\PassOptionsToPackage{caption=false}{subfig}
\usepackage{subfig}
\usepackage{multirow}
\usepackage{booktabs}
\usepackage{hyperref}

\begin{document}

\preprint{IPMU13-0140}
\preprint{UCB-PTH-13/06}

\title{A Natural Higgs Mass in Supersymmetry from Non-Decoupling Effects} 

\author{Xiaochuan Lu}
\email{luxiaochuan123456@berkeley.edu}
\affiliation{Department of Physics, University of California,
  Berkeley, California 94720, USA}  
\affiliation{Theoretical Physics Group, Lawrence Berkeley National
  Laboratory, Berkeley, California 94720, USA} 

\author{Hitoshi Murayama}
\email{hitoshi@berkeley.edu, hitoshi.murayama@ipmu.jp}
\affiliation{Department of Physics, University of California,
  Berkeley, California 94720, USA} 
\affiliation{Theoretical Physics Group, Lawrence Berkeley National
  Laboratory, Berkeley, California 94720, USA} 
\affiliation{Kavli Institute for the Physics and Mathematics of the
  Universe (WPI), Todai Institutes for Advanced Study, University of Tokyo,
  Kashiwa 277-8583, Japan} 

\author{Joshua T. Ruderman}
\email{ruderman@berkeley.edu}
\affiliation{Department of Physics, University of California,
  Berkeley, California 94720, USA} 
\affiliation{Theoretical Physics Group, Lawrence Berkeley National
  Laboratory, Berkeley, California 94720, USA} 

\author{Kohsaku Tobioka}
\email{kohsaku.tobioka@ipmu.jp, tobioka@post.kek.jp}
\affiliation{Kavli Institute for the Physics and Mathematics of the
  Universe (WPI), Todai Institutes for Advanced Study, University of Tokyo,
  Kashiwa 277-8583, Japan} 
\affiliation{Department of Physics, University of Tokyo,
  Hongo, Tokyo 113-0033, Japan}

\begin{abstract}
The Higgs mass implies fine-tuning for minimal theories of weak-scale supersymmetry (SUSY). Non-decoupling effects can boost the Higgs mass when new states interact with the Higgs, but new sources of SUSY breaking that accompany such extensions threaten naturalness.  We show that {two singlets} with a Dirac mass can increase the Higgs mass while maintaining  naturalness in the presence of large SUSY breaking in the singlet sector.  We explore the modified Higgs phenomenology of this scenario, which we call the ``Dirac NMSSM."
 \end{abstract}

\maketitle

{\bf Introduction: }The discovery of a new resonance at 125~GeV~\cite{:2012gk}, that appears to be the
long-sought Higgs boson, marks a great triumph of experimental and
theoretical physics.  On the other hand,  the presence of this light scalar forces us to face the naturalness problem of its mass.
Arguably, the best known mechanism
to ease the naturalness problem is weak-scale supersymmetry (SUSY), but the lack of experimental signatures is pushing 
SUSY into a tight corner.  In
addition, the observed mass of the Higgs boson 
is higher than what was expected in the Minimal Supersymmetric Standard
Model (MSSM), requiring fine-tuning of parameters at the 1\% level or worse~\cite{Hall:2011aa}.

If SUSY is realized in nature, one possibility is to give up on naturalness~\cite{Wells:2003tf}.  Alternatively, theories that retain naturalness must address two problems, (I) the missing superpartners and (II) the Higgs mass.  The collider limits on superpartners are highly model-dependent and can be relaxed when superpartners unnecessary for naturalness are taken to be heavy~\cite{Dimopoulos:1995mi}, when less missing energy is produced due to a compressed mass spectrum~\cite{LeCompte:2011fh} or due to decays to new states~\cite{Fan:2011yu}, and 
 when $R$-parity is violated~\cite{Barbier:2004ez}.  Even if superpartners have evaded detection for one of these reasons, we must address the surprisingly heavy Higgs mass.

There have been many attempts to extend the MSSM to accommodate the Higgs mass.  In such extensions, new states interact with the Higgs, raising its mass by increasing the strength of the quartic interaction of the scalar potential.  If the new states are integrated out supersymmetrically, their effects decouple and the Higgs mass is not increased.  On the other hand, SUSY breaking can lead to non-decoupling effects that increase the Higgs mass.  One possibility is a non-decoupling $F$-term, as in the  NMSSM (MSSM plus a singlet)~\cite{Espinosa:1991gr, Nomura:2005rk} or $\lambda$SUSY (allowing for a Landau
pole)~\cite{Harnik:2003rs}.  A second possibility is a non-decoupling $D$-term that results if the Higgs is charged under a new gauge group~\cite{Batra:2003nj}.   In general, these extensions require new states at the few hundred GeV scale, so that the new sources of SUSY breaking do not spoil naturalness.

\begin{figure}[b!]
\centering
\vspace{-10pt}
\includegraphics[width=0.6\linewidth]{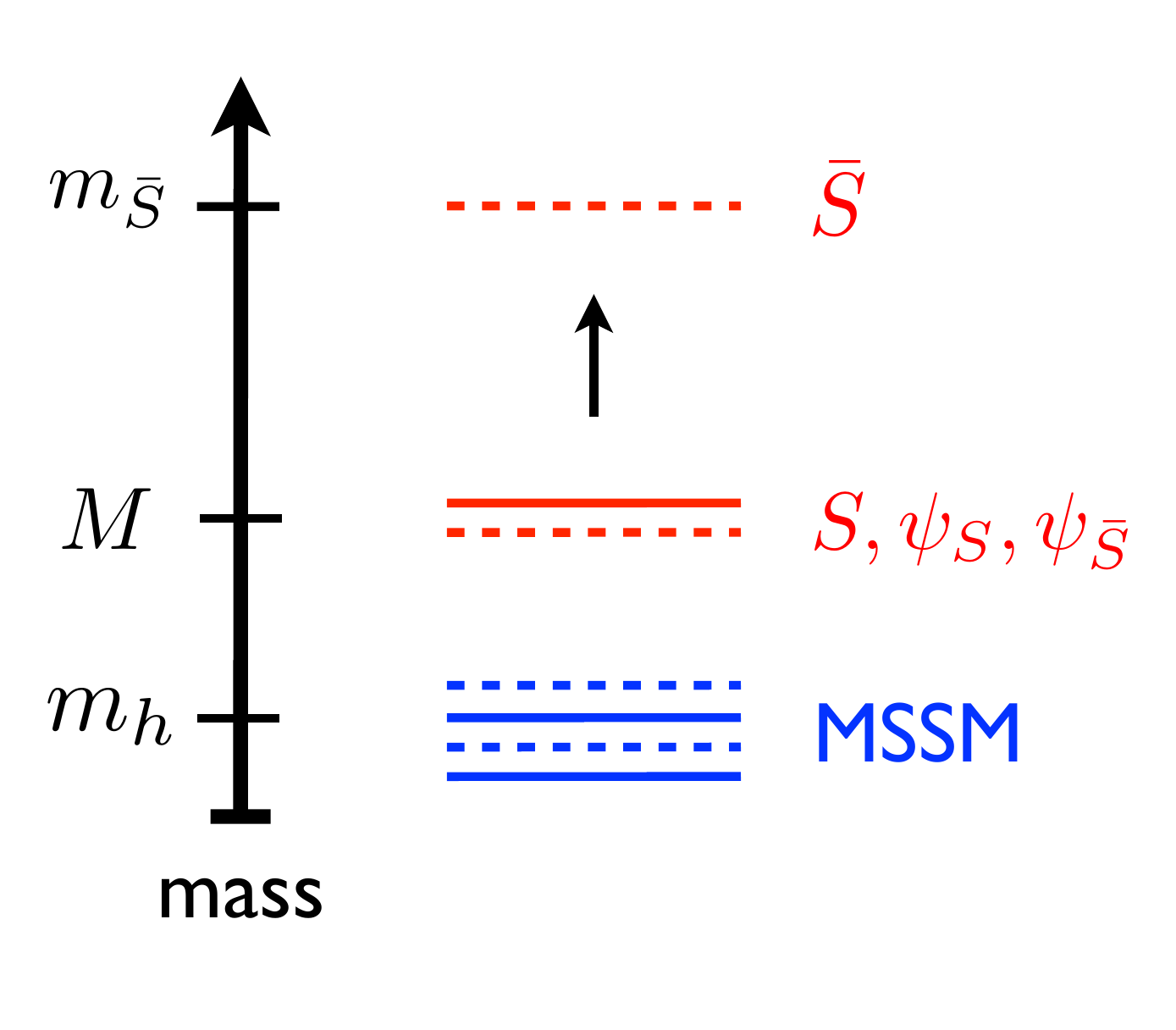}
\vspace{-10pt}
\caption{\label{fig:schema}
A typical spectrum of the Dirac NMSSM, which allows for large $m_{\bar S}$ without spoiling naturalness.
\\
}
\end{figure}

For example, consider the NMSSM, where a singlet superfield, $S$, interacts with the MSSM Higgses, $H_{u,d}$, through the superpotential,
\begin{equation}
  W \supset \lambda \, S H_u H_d + \frac{M}{2} S^2 +\mu \, H_u H_d.
  \label{eq:NMSSM}
\end{equation}
The Higgs mass is increased by,
\begin{equation} \label{eq:nmssm_mass}
\Delta m_h^2 = \lambda^2 v^2 \sin^2 2 \beta \left( \frac{m_S^2}{M^2 + m_S^2} \right),
\end{equation}
where $m_S^2$ is the SUSY breaking soft mass $m_S^2 |S|^2$, $\tan \beta = v_u / v_d$ is the ratio of the VEVs of the up and down-type Higgses,  and $v=\sqrt{v_u^2+v_d^2} = 174$~GeV\@.  Notice that this term decouples in the supersymmetric limit, $M \gg m_S$, which means $m_S$ should not be too small.  On the other hand, $m_S$ feeds into the Higgs soft masses, $m_{H_{u,d}}^2$ at one-loop, requiring fine-tuning if $m_S \gg m_h$.  
Therefore, with $M$ at the weak-scale, there is tension between raising the Higgs mass, which requires large $m_S$ relative to $M$, and naturalness, which demands small $m_S$.

In this letter, we point out that, contrary to the above example,  a {\it lack of light
  scalars}\/ can help raise the Higgs mass without a cost to
naturalness, if the singlet has a Dirac mass.  We begin by introducing the model and discussing the Higgs mass and naturalness properties.  Then, we discuss the phenomenology of the Higgs sector, which can be discovered or constrained with future collider data.  We finish with our conclusions.

\begin{figure*}[ht!]
\centering
\includegraphics[width=0.8\linewidth]{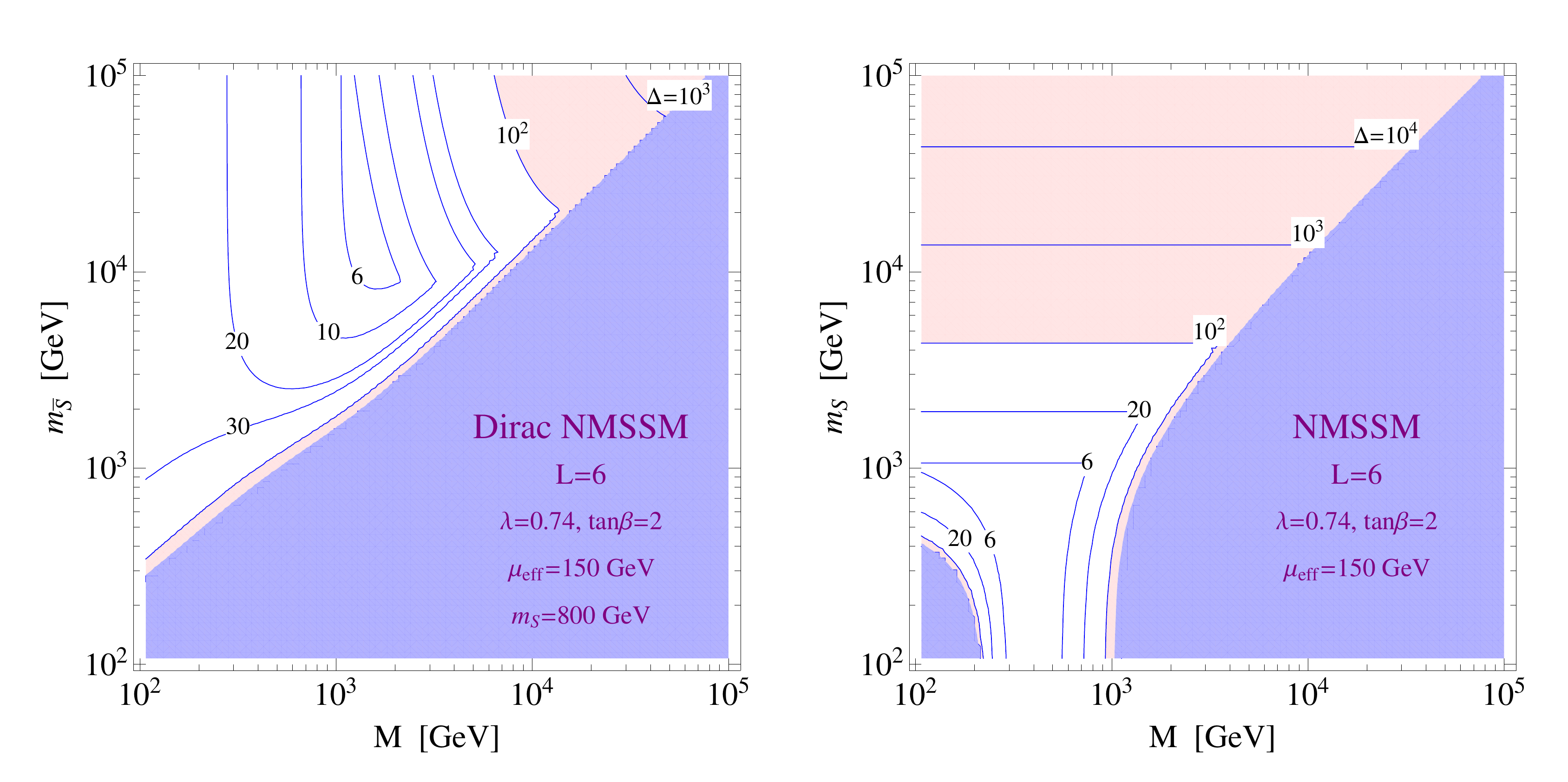}
\caption{\label{fig:tuning}
 The tuning $\Delta$, defined in Eq.~(\ref{eq:tune}), for the Dirac
 NMSSM is shown on the {\it left}\/ as a function of $M$ and $m_{\bar
   S}$.   For comparison, the tuning of the NMSSM is shown
 on the {\it right}\/, as a function of $M$ and
 $m_S$.  
 The red region has high fine-tuning, $\Delta >
 100$, and the purple region requires $m_{\tilde t} > 2$~TeV, signaling severe fine-tuning $\gtrsim \mathcal{O}(10^3)$.}
\vspace{-5pt}
\end{figure*}

{\bf The Model:}  To illustrate this possibility, we consider a modification of Eq.~(\ref{eq:NMSSM}) where $S$ receives a Dirac mass with another singlet, $\bar S$,
\begin{equation}
  W = \lambda \, S  H_u H_d + M S \bar{S} +\mu \, H_u H_d.
\end{equation}
We call this model the {\it Dirac NMSSM}.  The absence of various dangerous operators (such as large tadpoles for the singlets) follows from a $U(1)_{PQ}\times U(1)_{\bar{S}}$ Peccei-Quinn-like symmetry,
\begin{table}[ht.]
  \begin{tabular}{|c|   c  c c c c c |} \hline
      &$H_u$& $H_d$ & $S$ &$\bar{S}$ & $\mu$ & $M$  \\\hline
     $U(1)_{PQ}$ & $1$ & $1$ & $-2$ & $-2$ & $-2$ & $4$ 
    \\
     $U(1)_{\bar{S}}$& $0$& $0$& $0$ & $1$ &$0$  & $-1$
\\ \hline
  \end{tabular}
\end{table}\\
Here, $U(1)_{\bar{S}}$ has the effect of differentiating $S$ and $\bar
S$ and forbidding the operator $\bar S H_u H_d$.  Because $\mu$
  and $M$ explicitly break the $U(1)_{PQ}\times U(1)_{\bar{S}}$ symmetry, we
  regard them to be spurions originating from chiral superfields (``flavons'' \cite{ArkaniHamed:1996xm})
 so that holomorphy is used to avoid certain unwanted terms
   (``SUSY zeros'' \cite{Leurer:1993gy}).
  By classifying all possible operators induced by these spurions, we
  see that a 
tadpole for $\bar{S}$ is suppressed adequately,
\begin{eqnarray}
	W \supset c_{\bar{S}} \mu M \bar{S} \ ,
\end{eqnarray}
where $c_{\bar{S}}$ is a ${\cal O}(1)$ coefficient. 
Other terms involving only singlets are forbidden by the symmetries or suppressed by the cutoff.

The  following soft supersymmetry breaking terms are allowed by the symmetries,
\begin{eqnarray}
  \Delta V_{\it soft} &=& m_{H_u}^2 |H_u|^2+m_{H_d}^2 |H_d|^2+
  m_S^2 |S|^2 + m_{\bar{S}}^2 |\bar{S}|^2 
  \nonumber\\ 
  &&+  \lambda A_\lambda {S} H_u H_d +  M B_S S\bar{S} +  \mu B H_u H_d  +c.c.
  \nonumber\\ 
  &&+ t_{\bar{S}} \bar{S}+ t_S S+c.c.
\end{eqnarray}
The last tadpole arises from a non-holomorphic term $\mu^\dagger S$. 
{The small hierarchy, which we consider later, between the soft masses of $\bar{S}$ and others can be naturally obtained in gauge mediation models if $\bar{S}$ couples to the messengers in the superpotential.  It is easy to write down a model where $m^2_{\bar{S}}$ is positive at the one-loop level, while the soft masses for squarks and sleptons arise at the two-loop level via gauge mediation  \cite{Dvali:1996cu}.  The tadpole for $\bar{S}$ is generated at one-loop and hence $t_{\bar{S}} \simeq \mu M m_{\bar{S}} / 4\pi$, while soft masses and the tadpole for $S$ are generated at higher order.  
We checked these singlet tadpoles do not introduce extra tuning even with large $M$ and $m_{\bar{S}}$.}

We would like to understand whether the new quartic term, $|\lambda H_u H_d|^2$,  can naturally 
raise the Higgs mass. 
Integrating out $S$ and $\bar S$ we find the following potential for the doublet-like Higgses,
\begin{eqnarray}
  V_{\it eff}
  &=& |\lambda H_u H_d|^2 \left( 1 - \frac{M^2}{M^2 + m_{\bar{S}}^2}  \right)
  \label{eq:Veff}   
\\
    &-& \frac{\lambda^2 }
    {M^2 + m_{S}^2 }\left| A_\lambda H_u H_d+{\mu^*}(|H_u|^2+|H_d|^2) \right|^2
     .
   \nonumber
\end{eqnarray}
where we keep leading $(M^2 + m_{S, \bar{S}}^2)^{-1}$  terms and neglect the tadpole terms for simplicity.
The additional Higgs quartic term does not decouple when $m^2_{\bar{S}}$ is large, as in the NMSSM at large $m^2_{{S}}$.
The SM-like Higgs mass becomes,
\begin{eqnarray}
  m_h^2 &=& m_{h,{\rm MSSM}}^2(m_{\tilde{t}}) + \lambda^2 v^2 \sin^2 2\beta \left( \frac{m_{\bar{S}}^2}{M^2 + m_{\bar{S}}^2} \right)\nonumber \\
  & & -\frac{ \lambda^2 v^2 }{M^2 + m_S^2}\left|
    A_{\lambda} \sin 2\beta-2\mu^*
     \right|^2, 
     \label{eq:higgsmass}
\end{eqnarray}
in the limit where the VEVs and mass-eigenstates are aligned.

The Higgs sector is natural when there are no large radiative corrections to $m_{H_{u,d}}^2$.
The renormalization group (RG) of the up-type Higgs contains the terms,
\begin{eqnarray}
  \mu \frac{d}{d\mu} m_{H_u}^2 &=& \frac{1}{8\pi^2} 
   \left(3 y_t^2 [m_{\tilde{Q}_3}^2+m_{\tilde{t}_R}^2]+\lambda^2m_{S}^2 \right) + \ldots
  \label{eq:RGE}
\end{eqnarray}
While heavy stops or $m_{S}^2$ lead to fine-tuning, we find that $m_{\bar S}^2$ does not appear.
In fact, the RGs for  $m_{H_{u,d}}^2$ are independent of $m_{\bar S}^2$ to all orders in mass-independent schemes, because $\bar{S}$ couples to the MSSM+$S$ sector 
only through the  dimensionful coupling $M$.
There is logarithmic sensitivity to $m_{\bar{S}}^2$ from the one-loop finite
threshold correction,
\begin{eqnarray}
  \delta m_H^2 \equiv \delta m_{H_{u,d}}^2 
  = \frac{(\lambda M)^2}{(4\pi)^2} \log \frac{M^2 + m_{\bar{S}}^2}{M^2}\ .
  \label{eq:threshold}
\end{eqnarray}
which still allows for very heavy $m_{\bar{S}}^2$ without fine-tuning.

One may wonder if there are dangerous finite threshold corrections to $m_{H_u}^2$ at higher order, after integrating out $\bar S$.  In fact, there is no quadratic sensitivity to $m_{\bar S}^2$ to all orders.  This follows because any dependence on $m_{\bar S}^2$ must be proportional to $|M|^2$ (since $\bar S$ {becomes free} when $M \rightarrow 0$ and by conservation of $U(1)_{\bar S}$), but $ |M|^2 m_{\bar S}^2$ has too high mass dimension.  The mass dimension cannot be reduced from other mass parameters appearing in the denominator because threshold corrections are always analytic functions of IR mass parameters~\cite{Georgi:1994qn}.

It may seem contradictory that naturalness is maintained in the limit of very heavy $m_{\bar S}$, since removing the $\bar S$ scalar from the spectrum constitutes a hard breaking of SUSY\@.  The reason is that the effective theory, with the $\bar S$ fermion but no scalar present at low energies is actually  equivalent to a theory with only softly broken supersymmetry,
where the MSSM is augmented by the K\"{a}hler operators,
\begin{eqnarray}
\!\!\!\!\!\!\!\!\!\!\!\! {\cal K}_{\it eff} &= &\bar{S}^\dagger \bar{S}  \nonumber \\
&-& \theta^2\bar{\theta}^2 \left( M\, {\cal D}^\alpha  {S} \ {\cal D}_\alpha  \bar{S} + c.c. +M^2 | S + c_{\bar{S}} \mu|^2 \right)
  \label{eq:semisoft}
\end{eqnarray}
and where the scalar and $F$-term of $\bar{S}$ are reintroduced at low-energy but completely decoupled from the other states.
We call this mechanism {\it semi-soft supersymmetry breaking}. It is crucial that $\bar{S}$ couples to the other fields only through dimensionful couplings. 
Note that Dirac gauginos are a different example where adding new fields can lead to improved naturalness properties \cite{Fox:2002bu}.

The most natural region of parameter space, summarized in Fig.~\ref{fig:schema},
has $m_S$ and $M$ at the hundreds of GeV scale, to avoid large corrections to $m_{H_u}$, and large $m_{\bar S} \gtrsim 10$~TeV, to maximize the second term of Eq.~(\ref{eq:higgsmass}).  The tree-level contribution to the Higgs mass can be large enough such that $m_{\tilde{t}}$ takes a natural value at the hundreds of GeV scale.

We have performed a quantitative study of the fine-tuning in the Dirac NMSSM, shown to the left of Fig.~\ref{fig:tuning} as a function of $(M,m_{\bar S})$.
We computed the radiative corrections from the top sector to the Higgs mass
at RG-improved Leading-Log order, analogous to \cite{Carena:1995bx}.  We have confirmed that our results match the FeynHiggs software~\cite{Frank:2006yh}, for the MSSM, within $\Delta m_h\simeq1$ GeV in the parameter regime of interest.
We fix $A_t=0$ for simplicity, and other parameters are fixed according to the table, shown below.
Here, we adopt a parameter $\mu_{\it eff}\equiv \mu+\lambda \left<S\right>$ for convenience. 
We have chosen $\lambda$ to saturate the upper-limit such that it does not reach a Landau pole below the unification scale~\cite{Barbieri:2007tu}.
For each value of  $(M,m_{\bar S})$, the stop soft masses, $m_{\tilde{t} }=m_{\tilde{t}_R }=m_{\tilde{Q}_3 }$, are chosen to maintain the lightest scalar mass at $125$~GeV\@. 
The degree of fine-tuning is
estimated by 
\begin{equation} \label{eq:tune}
  \Delta = \frac{2}{m_h^2} {\rm max} \left( m_{H_u}^2, m_{H_d}^2, \frac{d m_{H_u}^2 }{d \ln \mu}
    L, \frac{d m_{H_d}^2}{d \ln \mu}  L, \delta m_H^2, b_{\it eff} \right),
\end{equation}
where $b_{\it eff}= \mu B+\lambda(A_\lambda \left<S\right>+M \left<{\bar{S}}\right>)$ and we take $L \equiv \log ( \Lambda / m_{\tilde t}) = 6$, corresponding to low-scale SUSY breaking.  
We assume that contributions through gauge couplings to the RGs for $m_{H_{u,d}}^2$ are subdominant.

\begin{table}[b!]
  \begin{tabular}{|c   c  c |}
\hline
 \multicolumn{3}{|c|}{benchmark parameters} \\ \hline
 $\lambda = 0.74$ & $\tan \beta = 2$ &\quad $\mu_{\it eff} = 150$~GeV \\
 $b_{\it eff}$ =(190 GeV)$^2$ &$A_\lambda = 0$ & $B_s = 100$~GeV\\ 
 $M = 1$~TeV  &\quad $m_{\bar S} = 10$~TeV& $m_S=800$~GeV\\
 \hline
  \end{tabular}
\end{table}

\begin{figure*}[t!]
\vspace{-15pt}
\centering
\includegraphics[width=1\linewidth]{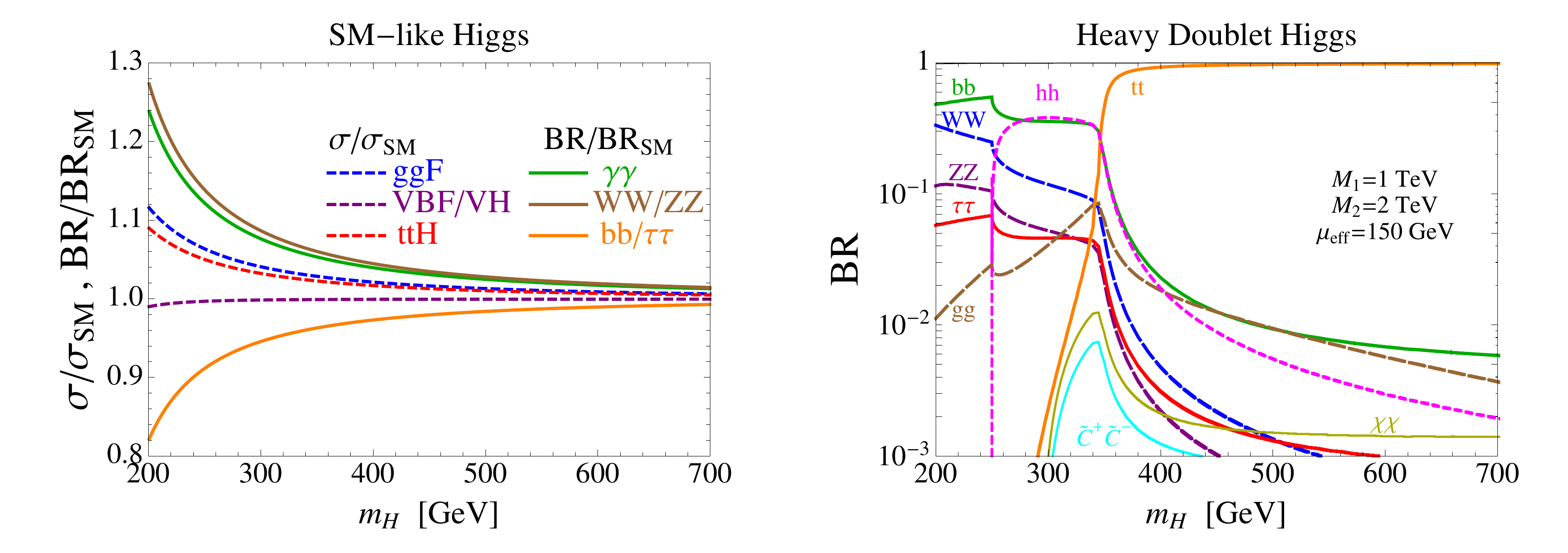}\vspace{-15pt}
\caption{\label{fig:DecayProduction}
The branching ratios and production cross sections of the SM-like Higgs are shown, normalized to the SM values~\cite{Dittmaier:2011ti, Dittmaier:2012vm}, on the {\it left} as a function of the heavy doublet-like Higgs mass, $m_H$.  On the {\it right}, we show several branching ratios of the heavy doublet-like Higgs as a function of its mass.  Note that the location of the chargino/neutralino thresholds depend on the -ino spectrum.  Here we take heavy gauginos and $\mu_{\it eff} = 150$~GeV.}
\end{figure*}
\begin{figure*}[t!]
\begin{center} \includegraphics[width=1\linewidth]{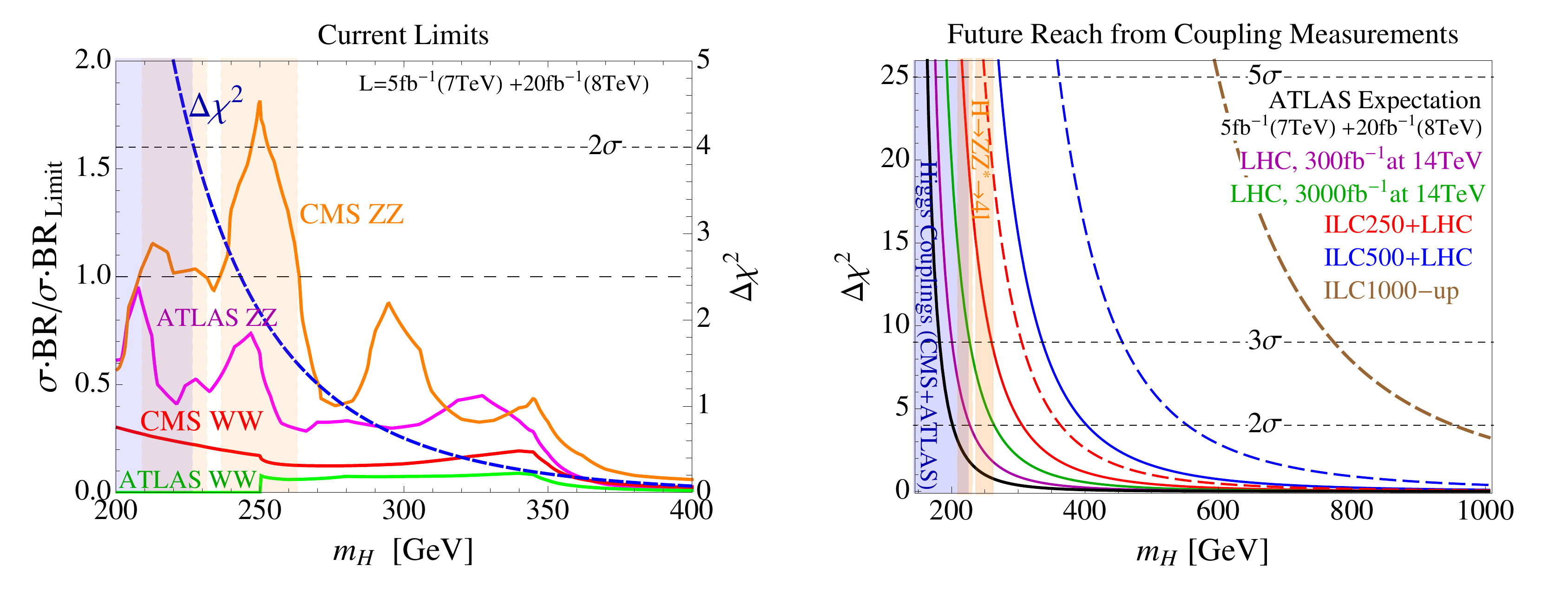} \end{center}
 \vspace{-15pt}
\caption{\label{fig:LimitReach}
The {\it left} plot shows current constraints on our model.  The right axis corresponds to $\Delta \chi^2$ for the SM-like Higgs couplings with the 7 and 8 TeV datasets \cite{ATLAS:2013sla, ATLASheavyHiggs, CMS:ril, CMSheavyHiggs}
neglecting correlations between the measurements of different couplings.  
 The left axis shows $\sigma / \sigma_{lim}$ for direct searches, $H \rightarrow ZZ,WW$~\cite{ATLASheavyHiggs, CMSheavyHiggs}. 
The {\it right} plot shows the expected $\Delta \chi^2$ from combined measurements of the Higgs-like couplings at the high-luminosity LHC at $\sqrt s = 14$~TeV~\cite{ATLAS-collaboration:2012iza} and the ILC at $\sqrt s = 250, 500, 1000$~GeV\@.  The optimistic~\cite{Peskin:2012we,Snowmass} (conservative~\cite{Klute:2013cx}) ILC reach curves are solid (dashed) and  neglect (include) theoretical uncertainties in the Higgs branching ratios. The ILC analyses include the expected LHC measurements.  For comparison we show the present limits and also the expected limit of the current ATLAS measurements (solid, black).}
\end{figure*}

For comparison, the right of Fig.~\ref{fig:tuning} shows the tuning in the NMSSM, which corresponds to the Dirac
NMSSM replacing $\bar{S} \rightarrow S$ (which removes the $U(1)_{\bar{S}}$ symmetry).
The superpotential of the NMSSM corresponds to  Eq.(\ref{eq:NMSSM}) plus the tadpole $c_S \mu M S$. We treat $m_S$ as a free parameter
instead of $m_{\bar{S}}$ and use the same fine-tuning measure of
Eq.(\ref{eq:tune}), except  the threshold correction $\delta m_H^2$ is absent and $b_{\it eff}= \mu B+\lambda \left<S\right>(A_\lambda +M)$.

We see that the least-tuned region of the Dirac NMSSM corresponds to $M \sim 2$~TeV and $m_{\bar S} \gtrsim 10$~TeV, where the tree-level correction to the Higgs mass is maximized. 
  The fine-tuning is dominated by  $\delta m_H^2$ in the large $M$ region, and by the contribution of $m_{\tilde{t}}$ to $m_{H_u}$ in the rest of the plane.  On the other-hand, the NMSSM becomes highly tuned when $m_S$ is large (since it radiatively corrects $m_{H_{u,d}})$, 
and then $m_S\lesssim 1~$TeV is favored. 
Note that region of low-tuning in the NMSSM extends to the supersymmetric limit, $m_{S} \rightarrow 0$.  In this region the Higgs mass is increased by a new contribution to the quartic coupling proportional to 
$\lambda^2(M\mu\sin 2\beta -\mu^2)/M^2$~(see Ref.\cite{Nomura:2005rk} for more details).

{\bf Higgs Phenomenology:}
We now discuss the experimental signatures of the Dirac NMSSM\@.  The phenomenology of the NMSSM is well-studied~\cite{Hall:2011aa, Ellwanger:2011aa}.  The natural region of the Dirac NMSSM differs from the NMSSM in that the singlet states are too heavy to be produced at the LHC\@.  The low-energy Higgs phenomenology is that of a two Higgs doublet model, and we focus here on the nature of the SM-like Higgs, $h$, and the heavier doublet-like Higgs, $H$ \cite{Craig:2012pu}.
The properties of the two doublets differ from the MSSM due to the presence of the non-decoupling quartic
coupling $|\lambda H_u H_d|^2$, which raises the Higgs mass by the semi-soft SUSY breaking, described above.  

We consider the potential with radiative corrections from the stop sector, and we find that the couplings of the SM-like Higgs to leptons and down-type quarks are lowered, while couplings to the up-type quarks are slightly increased compared to those in the SM, which results in the deviations to the cross sections and decay patterns shown to the left of Fig.~\ref{fig:DecayProduction}. These effects decouple in limit $m_H \gg m_h$, which corresponds to large $b_{\it eff}$. We also show, to the right of Fig.~\ref{fig:DecayProduction}, the decay branching ratios of $H$. Due to the non-decoupling term, di-Higgs decay, $H \rightarrow 2h$, becomes the dominant decay once its threshold is opened, $m_H\gtrsim 250$~GeV.

There are now two relevant constraints on the Higgs sector of the Dirac NMSSM\@.  The first comes from measurements of the couplings of the SM-like Higgs from ATLAS~\cite{ATLAS:2013sla, ATLASheavyHiggs} and CMS~\cite{CMS:ril, CMSheavyHiggs}.  The second comes from direct searches for the heavier state decaying to dibosons, $H \rightarrow ZZ,WW$~\cite{ATLASheavyHiggs, CMSheavyHiggs}.  The former excludes $m_H \lesssim 220$~GeV at $95\%$, while the latter extends this limit to $m_H \sim 260$~GeV {by the CMS  search for $H \rightarrow ZZ$} (except for a small gap near $m_H \approx 235$~GeV), as can be seen to the left of  Fig.~\ref{fig:LimitReach}.  
We also estimate the future reach to probe $m_H$ with future
Higgs coupling measurements
\cite{ATLAS-collaboration:2012iza, Peskin:2012we, Klute:2013cx, Snowmass}, shown to the right of Fig.~\ref{fig:LimitReach}. The $2 \sigma$ exclusion reach is
$m_H\simeq 280$~GeV at the high-luminosity LHC \cite{ATLAS-collaboration:2012iza},
$m_H\simeq 400$~GeV with theoretical uncertainty at ILC500 \cite{Klute:2013cx},
and $m_H\simeq 950$~GeV without theoretical uncertainty at upgraded ILC1000 \cite{Snowmass}.
The increased sensitivity at the ILC is dominated by the improved measurements projected for the $b\bar{b}$ and $\tau^+\tau^- $ couplings~\cite{Peskin:2012we,Snowmass}.

{\bf Discussion:} 
The LHC has discovered a new particle, consistent with the Higgs boson, with a mass near 125~GeV\@.  Weak-scale SUSY must be reevaluated in light of this discovery.
Naturalness demands new dynamics beyond the minimal theory, such as a non-decoupling $F$-term, but this implies new sources of SUSY breaking that themselves threaten naturalness.  In this paper, we have identified a new model where the Higgs couples to a singlet field with a Dirac mass.  The non-decoupling $F$-term is naturally realized through semi-soft SUSY breaking, because large $m_{\bar S}$ helps raise the Higgs mass but does not threaten naturalness.  The first collider signatures of the Dirac NMSSM are expected to be those of the MSSM fields, with the singlet sector naturally heavier than 1 TeV.

The key feature of semi-soft SUSY breaking in the Dirac NMSSM is that $\bar S$ couples to the MSSM only through the dimensionful Dirac mass, $M$.   We note that interactions between $\bar S$ and other new states are not constrained by naturalness, even if these states experience SUSY breaking.  Therefore, the Dirac NMSSM represents a new type of portal, whereby our sector can interact with new sectors, with large SUSY breaking, without spoiling naturalness in our sector.

{\it\bf Acknowledgments:} 
We thank Lawrence Hall, Matthew McCullough, Satyanarayan Mukhopadhyay, Tilman Plehn, Filippo Sala, Satoshi Shirai, and Neal Weiner for helpful discussions.  We especially thank Yasunori Nomura for discussions and for pointing out that the Dirac mass can be thought of as a new type of portal.
The work of HM was supported in part by the U.S. DOE under Contract No.~DEAC03-76SF00098, by the NSF under Grant No.~PHY-1002399, by the JSPS Grant (C) No.~23540289, by the
FIRST program Subaru Measurements of Images and
Redshifts (SuMIRe), CSTP, and by WPI, MEXT, Japan.
JTR is supported by a fellowship from the Miller Institute for
Basic Research in Science. The work of KT is supported in part by the
Grant-in-Aid for JSPS Fellows.  

\end{document}